\documentclass[amssymb,prb,floats,aps,superscriptaddress,twocolumn,narrowtext]{revtex4}
\usepackage{amsmath}
\usepackage{exscale}
\usepackage[mathscr]{eucal}
\usepackage{epsfig}

\begin{document}

\title{Nernst effect and the thermoelectric power in
strongly coupled electron-phonon systems}

\author{K. K. Lee}
\affiliation{IRC in Superconductivity, Cavendish Laboratory,
University of Cambridge, Cambridge, CB3 0HE, United Kingdom}

\begin{abstract}
An understanding of high temperature superconductivity clearly
requires the nature of its normal state, however the normal state
of the cuprates is poorly understood. One possible explanation is
the bipolaronic model where mobile, heavily phonon-dressed holes
(polarons) are paired in real space. The ground state is made up
of intersite singlet bipolarons where excitations can be made to
triplet bipolaron and polaron bands. Based on this model I derive
the Nernst coefficient and the Thermoelectric power for the
cuprates. The Nernst coefficient shows an interesting interference
term due to the triplet and singlet bipolaron and polaron current
flow. The interference remains even when the scattering rate is
independent of energy.
\end{abstract}

\vskip2pc

\narrowtext

\bigskip
\maketitle

\section{Introduction}
To account for the high values of T$_{c}$ in the cuprates, it is
necessary to have electron-phonon interactions larger than those
found in the intermediate coupling theory of
superconductivity\cite{ELIA}. Regardless of the adiabatic ratio,
the Migdal-Eliashberg theory of superconductivity and
Fermi-liquids has been shown to breakdown at $\lambda =1$
\cite{ALEX} using the ($1/\lambda $) expansion
technique\cite{LANG}. The many-electron system collapses into the
small polaron regime\cite{ALEX,SASH,SAS} at $\lambda \geq 1$ with
well separated vibration and charge-carrier degrees of freedom.
Moreover the electron-phonon interaction is sufficiently large to
bind small polarons into small bipolarons. At first sight these
carriers have a mass too large to be mobile, however it has been
shown that the inclusion of the on-site coulomb repulsion leads to
the favoured binding of intersite oxygen holes\cite{ALEXAND,
SANF}. The intersite bipolarons can then tunnel with an effective
mass of only 10 electron masses\cite{ALEXAND,CATL,alekor,tru}. In
the cuprates bipolarons are real-space pairs\cite{alexcond} due to
the small fermi energy, where the condition for real-space pairing
is
\begin{equation}
\epsilon _{F}\lesssim \pi \Delta.
\end{equation}
Here $\Delta$ is the binding energy of a bipolaron. Mott and
Alexandrov proposed a simple model\cite{MOTT} of the cuprates
based on bipolarons. In this model all the holes (polarons) are
bound into small intersite singlet and triplet bipolarons at any
temperature. Above T$_{c}$ this is a non-degenerate charged bose
gas and below T$_{c}$ we have a Bose-Einstein condensate. The
triplet and singlet states are separated by an exchange energy J
which explains the spin gap observed in many NMR and neutron
scattering experiments\cite{ROSS,MOOK}. Also the bipolaron binding
energy has been suggested to be twice the so-called pseudogap
\cite{SANF}.

Transport measurements are vital in understanding the nature of
the carriers and so here I present two kinetic coefficients which
help describe the characteristics of a bipolaronic gas.

\section{Electrical transport}
The standard Boltzmann equation for kinetics was applied to well
decoupled, renormalised carriers\cite{kkl}. In the presence of the
electric field {\bf E}, temperature gradient ${\bf \nabla }{T}$
and magnetic field {\bf B}$\parallel $ {\bf z} $\perp $ {\bf E}
and ${\bf \nabla }{T}$, the electrical current for each carrier is
given by
\begin{equation}
{\bf j}^{\alpha }{\bf =}q\sum_{{\bf k}}{\bf v}f_{1}^{\alpha }({\bf
k})
\end{equation}
where $\alpha =s,p,t$, and $q$ is the carrier charge. We find the
x direction component as
\begin{eqnarray}
j_{x}^{\alpha } &=&a_{xx}^{\alpha }\nabla _{x}(\mu -2e\phi
)+a_{xy}^{\alpha
}\nabla _{y}(\mu -2e\phi )  \nonumber \\
&&+b_{xx}^{\alpha }\nabla _{x}T+b_{xy}^{\alpha }\nabla _{y}T
\end{eqnarray}
and also the y direction component
\begin{eqnarray}
j_{y}^{\alpha } &=&a_{yy}^{\alpha }\nabla _{y}(\mu -2e\phi
)+a_{yx}^{\alpha
}\nabla _{x}(\mu -2e\phi )  \nonumber \\
&&+b_{yy}^{\alpha }\nabla _{y}T+b_{yx}^{\alpha }\nabla _{x}T
\end{eqnarray}
where
\begin{eqnarray}
a_{xx}^{p} &=&a_{yy}^{p}=\frac{en_{p}}{2m_{p}}\langle \tau _{p}\rangle , \\
a_{yx}^{p} &=&-a_{xy}^{p}=\frac{eg_{p}Bn_{p}}{2m_{p}}\langle \tau
_{p}^{2}\rangle ,  \nonumber \\
b_{xx}^{p} &=&b_{yy}^{p}=\frac{en_{p}}{Tm_{p}}\langle \tau
_{p}\{E+\Delta
/2-\mu /2\}\rangle ,  \nonumber \\
b_{yx}^{p} &=&-b_{xy}^{p}=\frac{eg_{p}Bn_{p}}{Tm_{p}}\langle \tau
_{p}^{2}\{E+\Delta /2-\mu /2\}\rangle ,  \nonumber \\
a_{xx}^{s,t} &=&a_{yy}^{s,t}=\frac{2en_{s,t}}{m_{s,t}}\langle \tau
_{s,t}\rangle ,  \nonumber
\\
a_{yx}^{s,t}
&=&-a_{xy}^{s,t}=\frac{2eg_{s,t}Bn_{s,t}}{m_{s,t}}\langle \tau
_{s,t}^{2}\rangle ,  \nonumber
\\
b_{xx}^{s} &=&b_{yy}^{s}=\frac{2en_{s}}{Tm_{s}}\langle \tau
_{s}\{E-\mu
\}\rangle ,  \nonumber \\
b_{yx}^{s} &=&-b_{xy}^{s}=\frac{2eg_{s}Bn_{s}}{Tm_{s}}\langle \tau
_{s}^{2}\{E-\mu \}\rangle  \nonumber
\\
b_{xx}^{t} &=&b_{yy}^{t}=\frac{2en_{t}}{Tm_{t}}\langle \tau
_{t}\{E+J-\mu
\}\rangle ,  \nonumber
\end{eqnarray}
\begin{eqnarray}
b_{yx}^{t} &=&-b_{xy}^{t}=\frac{2eg_{t}Bn_{t}}{Tm_{t}}\langle \tau
_{t}^{2}\{E+J-\mu \}\rangle ,  \nonumber
\end{eqnarray}
and
\begin{equation}
\langle \tau _{\alpha }^{r}\rangle =\frac{\int_{0}^{\infty
}dEE\tau _{\alpha }^{r}(E)[1+(g_{\alpha }\tau _{\alpha
}(E)B)^{2}]^{-1}\partial f_{0}^{\alpha }/\partial
E}{\int_{0}^{\infty }dEf_{0}^{\alpha }}
\end{equation}

Here $\mu $ is the chemical potential, $J$ is the exchange energy
which separates the triplet state from the singlet state, and
$\Delta $ is the bipolaron binding energy per pair which is
assumed to be of s-symmetry. The latter assumption of s-wave bulk
pairing symmetry of a single-particle gap has been shown to be a
valid one \cite{mul2,AND,ZHAO1}. Also we have set $\hbar =1$ and
also from now on set $k_{B}=c=1$. ${\bf v} =\partial E/\partial
{\bf k}$, $\tau $ is the relaxation time and we assume that it
depends on the kinetic energy, $E=k^{2}/2m$.  $g=g_{s}=2e/m_{s}$
for singlet bipolarons with the energy $E=k^{2}/(2m_{s})$. For
triplet bipolarons, $g=g_{t}=2e/m_{t}$ and the energy
$E=k^{2}/(2m_{t})$. Similarly for thermally excited polarons
$E=k^{2}/(2m_{p})$ and $g=g_{p}=e/m_{p}$ . Here $m_{s,t,p}$ are
the singlet and triplet bipolaron and polaron masses of
$two$-dimensional carriers.

The number densities, $n_{\alpha }$ of the three carriers can be
evaluated as
\begin{equation}
n_{p}=\frac{m_{p}T}{\pi }\ln \left[ 1+\exp \left( -\frac{\Delta
-\mu }{2T}\right) \right] ,  \label{29}
\end{equation}
\begin{equation}
n_{s}=-\frac{m_{s}T}{2\pi }\ln \left[ 1-\exp \left( \frac{\mu
}{T}\right)\right] ,  \label{30}
\end{equation}
\begin{equation}
n_{t}=-\frac{3m_{t}T}{2\pi }\ln \left[ 1-\exp \left( \frac{\mu
-J}{T}\right)\right] .  \label{31}
\end{equation}

Both the kinetic coefficients considered in this paper are caused
by an applied thermal gradient rather than an applied potential
gradient. Also they both occur in the absence of an electrical
current hence ${\bf j}={\bf j_{p}}+{\bf j_{s}}+{\bf j_{t}}=0$.
Using eqns. (3) and (4), we find,
\begin{eqnarray}
\nabla_x(\mu-2e\phi)=-\frac{b_{yx}a_{yx}+b_{xx}a_{xx}}{(a_{xx})^2+(a_{yx})^2}\nabla_xT
\nonumber\\
+\frac{b_{yx}a_{xx}-b_{xx}a_{yx}}{(a_{xx})^2+(a_{yx})^2}\nabla_yT
\end{eqnarray}
and
\begin{eqnarray}
\nabla_y(\mu-2e\phi)=-\frac{b_{yy}a_{yy}+b_{yx}a_{yx}}{(a_{xx})^2+(a_{yx})^2}\nabla_yT
\nonumber\\
+\frac{b_{xx}a_{yx}-b_{yx}a_{xx}}{(a_{xx})^2+(a_{yx})^2}\nabla_xT
\end{eqnarray}
where $b_{xx}=b_{xx}^p+b_{xx}^s+b_{xx}^t$,
$a_{yx}=a_{yx}^p+a_{yx}^s+a_{yx}^t$, etc.

These two equations can used to eliminate the potential gradient
terms from other kinetic equations involving only the temperature
gradient as the non-equilibrium source.

\section{Nernst coefficient - Strong and Weak field}
The Nernst effect, $Q$ is similar to the Hall effect except here
the induced Hall field is created by a thermal gradient. There is
no applied potential gradient and the carriers are affected by a
thermal gradient only ($\parallel $ {\bf xy}). The charged
carriers are then deflected perpendicular to the charge flow by an
applied magnetic field ($\parallel $ {\bf z}) setting up the
Nernst electric field, $E_y=-QB{\bf \nabla}_xT$. Using eqn(11) we
obtain the isothermal Nernst coefficient (${\bf \nabla}_yT=0$)
\begin{equation}
Q=\frac{b_{yx}a_{xx}-b_{xx}a_{yx}}{2eB(a_{xx}^2+a_{xy}^2)}
\end{equation}
If the magnetic field is weak, $g_{\alpha }\tau _{\alpha }B<<1,$
we can ignore all terms in $B^{2}$ and higher order. The previous
definition of $\langle \tau _{\alpha }^{r}\rangle $ becomes
\begin{equation}
\langle \tau _{\alpha }^{r}\rangle =\frac{\int_{0}^{\infty }E\tau
_\alpha^{r}dE\partial f_{0}^{\alpha }/\partial E}{\int_{0}^{\infty
}dEf_{0}^{\alpha }}
\end{equation}

\begin{eqnarray}
Q^{weak}=[\frac{g_p<\tau_p^2>D_1^2}{eT<\tau_p>}][(\Sigma_p-\Gamma_p)
\\
+2A_{s2}(\Sigma_s-2\Gamma_p-\Delta)+8A_{t1}A_{s2}(\Sigma_s-\Gamma_t-J)]\nonumber
\\
+8A_{s1}A_{s2}(\Sigma_s-\Gamma_s)+2A_{t2}(\Sigma_t-2\Gamma_p+J-\Delta)]\nonumber
\\
+2A_{s1}(2\Sigma_p-\Gamma_s+\Delta)+8A_{s1}A_{t2}(\Sigma_t+J-\Gamma_s)\nonumber
\\
+2A_{t1}(2\Sigma_p+\Delta-J-\Gamma_t)+8A_{t1}A_{t2}(\Sigma_t-\Gamma_t)\nonumber
\end{eqnarray}
where
\begin{equation}
A_{s1,t1}=\frac{n_{s,t}\langle \tau _{s,t}\rangle
m_{p}}{n_{p}\langle \tau _{p}\rangle m_{s,t}}.  \label{21}
\end{equation}
\begin{equation}
A_{s2,t2}=\frac{g_{s,t}n_{s,t}\langle \tau _{s,t}^{2}\rangle m_{p}}{%
g_{p}n_{p}\langle \tau _{p}^{2}\rangle m_{s,t}}.  \label{22}
\end{equation}

and have introduced the parameters
\begin{equation}
\Sigma _{\alpha}=\frac{\int_{0}^{\infty }dEE^{2}\tau _{\alpha
}^2(E)\partial f_{0}^{\alpha }/\partial E}{\int_{0}^{\infty
}dEE\tau _{\alpha }^2(E)\partial f_{0}^{\alpha }/\partial E},
\label{10}
\end{equation}
\begin{equation}
\Gamma _{\alpha }=\frac{\int_{0}^{\infty }dEE^{2}\tau _{\alpha
}(E)\partial f_{0}^{\alpha }/\partial E}{\int_{0}^{\infty }dEE\tau
_{\alpha }(E)\partial f_{0}^{\alpha }/\partial E}, \label{10}
\end{equation}
and $D_{1,2}=(1+4A_{s1,2}+4A_{t1,2})^{-1}$.

Here the $\Gamma$ terms are representative of a type of scattering
mechanism(s), which can be a combination of types. In the
non-degenerate system (above $T_c$) it is given by
\begin{equation}
\Gamma =(r+2)  \label{33}
\end{equation}
where $r$ is related to the energy dependence of the scattering
time
\begin{equation}
\tau \propto E^{r}.  \label{34}
\end{equation}

If the magnetic field is strong, $g_{\alpha }\tau _{\alpha }B>>1,$
we can gather all terms in $B^{2}$ and ignore the lower order
terms.

\begin{eqnarray}
Q^{strong}=[\frac{<\tau_p>D_2^2}{eTg_pB^2<\tau_p^2>}][(\Sigma_p-\Gamma_p)
\\
+2A_{s2}(\Sigma_s-2\Gamma_p-\Delta)+8A_{t1}A_{s2}(\Sigma_s-\Gamma_t-J)]\nonumber
\\
+8A_{s1}A_{s2}(\Sigma_s-\Gamma_s)+2A_{t2}(\Sigma_t-2\Gamma_p+J-\Delta)]\nonumber
\\
+2A_{s1}(2\Sigma_p-\Gamma_s+\Delta)+8A_{s1}A_{t2}(\Sigma_t+J-\Gamma_s)\nonumber
\\
+2A_{t1}(2\Sigma_p+\Delta-J-\Gamma_t)+8A_{t1}A_{t2}(\Sigma_t-\Gamma_t)\nonumber
\end{eqnarray}
and now the definition of $\langle \tau _{\alpha }^{r}\rangle $
becomes
\begin{equation}
\langle \tau_{\alpha }^{r}\rangle =\frac{\int_{0}^{\infty
}\frac{E\tau_{\alpha }^{r}}{(g_\alpha B\tau _\alpha)^2}dE\partial
f_{0}^{\alpha }/\partial E}{\int_{0}^{\infty }dEf_{0}^{\alpha }}
\end{equation}

In overdoped and optimally doped cuprates the exchange energy
between the triplet and singlet states, J $\sim 0$. Therefore we
may consider a degenerate singlet/triplet bipolaron system in this
regime where we now have
\begin{eqnarray}
Q^{weak}=[\frac{g_p<\tau_p^2>D_1^2}{eT<\tau_p>}][(\Sigma_p-\Gamma_p)
\\
+8A_{b1}A_{b2}(\Sigma_b-\Gamma_b)\nonumber
\\
+2A_{b2}(\Sigma_b-2\Gamma_p-\Delta)\nonumber
\\
+2A_{b1}(2\Sigma_p-\Gamma_b+\Delta)]\nonumber
\end{eqnarray}
and
\begin{eqnarray}
Q^{strong}=[\frac{<\tau_p>D_2^2}{eTg_pB^2<\tau_p^2>}][(\Sigma_p-\Gamma_p)
\\
+8A_{b1}A_{b2}(\Sigma_b-\Gamma_b)\nonumber
\\
+2A_{b2}(\Sigma_b-2\Gamma_p-\Delta)\nonumber
\\
+2A_{b1}(2\Sigma_p-\Gamma_b+\Delta)]\nonumber
\end{eqnarray}

where now $D_{1,2}=(1+4A_{b1,2})^{-1}$ and the bipolaron number
density becomes
\begin{equation}
n_{b}=-\frac{2m_{b}T}{\pi }\ln \left[ 1-\exp \left( \frac{\mu
}{T}\right)\right] ,  \label{30}
\end{equation}

It is interesting to note that if $\tau_\alpha(E)=\tau_\alpha$ for
all the carriers, the Nernst coefficient remains finite for both a
degenerate and non-degenerate singlet/triplet system. When the
scattering rate is independent of energy, $\Sigma=\Gamma$. Hence
for a single carrier system the Nernst coefficient is zero and so
the presence of different carriers leads to interference terms.
Also the Nernst coefficient in a strong magnetic field yields the
expected $1/B^2$ dependence.

\section{Thermoelectric power}
When an electric field is induced parallel to the applied thermal
gradient in the absence of an electrical current (with no applied
potential gradient or magnetic field), we can define the
Thermoelectric power, $\alpha$ by
\begin{equation}
E_x=\alpha{\bf \nabla}_xT
\end{equation}

We can use either eqn(10) or (11) omitting the off diagonal terms
giving
\begin{eqnarray}
\alpha=-\frac{(\Gamma_p+(\Delta-\mu)/2)+2A_{s1}(\Gamma_s-\mu)}{eT(1+4A_{s1}+4A_{t1})}
\\
+\frac{2A_{t1}(\Gamma_t+J-\mu)}{eT(1+4A_{s1}+4A_{t1})} \nonumber
\end{eqnarray}
Again for a degenerate triplet/singlet system we have
\begin{equation}
\alpha=-\frac{(\Gamma_p+(\Delta-\mu)/2)+2A_{b1}(\Gamma_b-\mu)}{eT(1+4A_{b1})}
\end{equation}
When comparing these formulae to experiment one needn't take into
account the phonon drag effect. The effect would usually occur
because the temperature gradient also causes phonons to flow from
the hot end to the cold end of the sample. This causes the
carriers to be "dragged" along by the phonons. However there is no
polaron-phonon interaction since it has been removed by the
Lang-Firsov canonical transformation\cite{SANF}. The dominant
scattering mechanism should be the polaron-polaron scattering.

\section{Conclusions}
Both the Nernst coefficient and Thermoelectric power have been
derived for a strongly coupled electron-phonon system with the
cuprates in mind. A triplet and singlet bipolaron and polaron
system have been considered as well as a system in which the
singlet/triplet are degenerate. Interestingly the Nernst
coefficient shows an intereference term even with an energy
independent scattering rate in both cases. A direct comparison
with the cuprates awaits reliable experimental data.

This work was supported by the EPSRC UK (grant R46977). I would
like to thank A.S. Alexandrov and W.Y. Liang for their helpful
comments.

\end{document}